\documentclass[aps,pra,preprint,nofootinbib,superscriptaddress]{revtex4-1}
\usepackage{subfigure,graphicx} 
\usepackage{amsfonts,amssymb,amsmath}
\usepackage{flafter}
\usepackage{multirow}
\usepackage{txfonts}
\usepackage[unicode,breaklinks]{hyperref}
\usepackage{siunitx}
\newcommand{\papertitle}{Simulating cosmological supercooling with a cold atom system II}
\hypersetup{
    unicode=true,
    a4paper=true,
    plainpages=false,
    pdftitle={\papertitle},
    pdfauthor={Thomas P. Billam, Kate Brown, Andrew J. Groszek, Ian G. Moss},
    pdfsubject={\papertitle},
    colorlinks=true,
    linkcolor=blue,
    citecolor=blue,
    filecolor=black,
    urlcolor=blue
}

\usepackage{ulem}

\newcommand{\be}{\begin{equation}}
\newcommand{\ee}{\end{equation}}
\newcommand{\bea}{\begin{eqnarray}}
\newcommand{\eea}{\end{eqnarray}}
\newcommand{\beal}{\begin{aligned}}
  \newcommand{\eeal}{\end{aligned}}

\newcommand{\eqnrefp}[1]{{[Eq.~(\ref{#1})]}}

\newcommand{\figreft}[2]{Fig.~\ref{#1}#2}

\begin{document} 

\title{\papertitle}

\author{Thomas P.\ Billam}
\email{thomas.billam@ncl.ac.uk}
\affiliation{Joint Quantum Centre (JQC) Durham--Newcastle, School of Mathematics, Statistics and Physics, 
Newcastle University, Newcastle upon Tyne, NE1 7RU, UK}

\author{Kate Brown}
\email{k.brown@ncl.ac.uk}
\affiliation{School of Mathematics, Statistics and Physics, 
Newcastle University, Newcastle upon Tyne, NE1 7RU, UK}

\author{Andrew J. Groszek}
\email{a.groszek@uq.edu.au}
\affiliation{Joint Quantum Centre (JQC) Durham--Newcastle, School of Mathematics, Statistics and Physics, 
Newcastle University, Newcastle upon Tyne, NE1 7RU, UK}
\affiliation{ARC Centre of Excellence in Future Low-Energy Electronics 
Technologies, School of Mathematics and Physics, University of Queensland, 
St Lucia, QLD 4072, Australia}
\affiliation{ARC Centre of Excellence for Engineered Quantum Systems, 
School of Mathematics and Physics, University of Queensland, St Lucia, 
QLD 4072, Australia}

\author{Ian G. Moss}
\email{ian.moss@ncl.ac.uk}
\affiliation{School of Mathematics, Statistics and Physics, 
Newcastle University, Newcastle upon Tyne, NE1 7RU, UK}

\date{\today}

\begin{abstract}
We perform an analysis of the supercooled state in an analogue of an 
early universe phase transition based on a one dimensional, two-component 
Bose gas with time-dependent interactions. We demonstrate that the system
behaves in the same way as a thermal, relativistic Bose gas undergoing
a first order phase transition. We propose a way to prepare the state of the system in the metastable phase as 
an analogue to supercooling in the early universe. While we show that parametric resonances in the system can
be suppressed by thermal damping, we find that the theoretically estimated
thermal damping in our model is too weak to suppress the resonances for
realistic experimental parameters. However, we propose that experiments to
investigate the effective damping rate in experiments would be worthwhile.
\end{abstract}

\maketitle

\section{Introduction}

Recent experiments using (quasi-)one-dimensional Bose gases
\cite{LangenExperimental2015, ErneUniversal2018, PruferObservation2018} have
demonstrated the potential of ultracold atom systems to be used as test-bed
systems with which to study many-body quantum dynamics. Such systems are a key
candidate for quantum simulators of cosmological
processes~\cite{RogerNatCom16,Eckel:2017uqx}.

The present work is related to simulating phase transitions in the very early
universe. In some of these phase transitions, the universe would have
supercooled into a metastable phase, or even into a `false vacuum' state, before
undergoing a first order phase transition. The ensuing violent fluctuations in
density would have echoes in the present day universe in the form of signals in
the cosmic microwave background \cite{PhysRevD.84.043507} and in a background of
gravitational waves \cite{Caprini:2009fx, Hindmarsh:2013xza}.

The simulation of false vacuum decay in an ultracold atom experiment
has already been discussed~\cite{FialkoFate2015,
  FialkoUniverse2017, Braden:2017add, Braden:2018tky, BillamSimulating2019}.
The scheme of Fialko et al. \cite{FialkoFate2015,FialkoUniverse2017}
uses a two-component Bose gas formed from two spin states of a spinor
condensate, coupled by a time-modulated microwave field. After time-averaging,
one obtains an effective description containing a metastable false vacuum state
in addition to the true vacuum ground state. Within this description the
relative phase between the two components behaves like a relativistic scalar
field, making this an ideal system for reproducing features of high energy
particle physics.

Refs.~\cite{FialkoFate2015,FialkoUniverse2017,Braden:2017add} studied the
decay of the false vacuum using field-theoretical instanton techniques
\cite{Coleman:1977py,Callan:1977pt}
and numerical simulations using the truncated Wigner technique~\cite{Steel1998,blakie_dynamics_2008}. 
However, Refs.~\cite{Braden:2017add,Braden:2019vsw} showed that the
false vacuum state in this scheme can suffer from a parametric instability
caused by the time-modulation of the system. The instability causes decay of the
false vacuum state by a different mechanism than a first-order phase transition.
This instability presents a challenge to experimental implementation of the
scheme.

This scheme was recently extended to a finite-temperature 1D Bose gas, with
temperatures in the phase-fluctuating quasi-condensate
regime~\cite{Billam:2020xna}, with the aim of studying thermodynamical first
order phase transitions in a cold atom system. Working in the time-averaged
effective description, both instanton techniques and the stochastic projected
Gross--Pitaevskii equation (SPGPE) \cite{GardinerStochastic2002, GardinerStochastic2003,
  bradley_bose-einstein_2008, BradleyStochastic2014} were used to
investigate the decay of a supercooled gas which has been prepared in the
metastable state at low (but nonzero) temperatures. These methods showed excellent agreement
in their predictions for the rate of the resulting first-order phase transition.
Furthermore, the results hold out the possibility that the thermal dissipation
present in the system could potentially stabilize the system against the
parametric instability. However, because the time-averaged effective description
was used, the mechanism of the parametric instability was explicitly removed in
the model of Ref.~\cite{Billam:2020xna}. To understand whether experiments could
realistically observe phase transition dynamics from a supercooled metastable
state using this scheme, two key questions to answer are: (i) how to prepare the
ultracold atom system in the appropriate metastable state, and (ii) whether
supercooled metastable states can be stabilized against the parametric
instability for realistic levels of dissipation encountered in a
finite-temperature Bose gas experiment.

In this paper, we address these questions using an SPGPE description of the full
time-dependent system. With regard to metastable state preparation, we propose
that the system is initialized first in a stable state, and then the
microwave field is adjusted by a control parameter in
such a way that the stable state becomes metastable. This is analagous to what
happens in the early universe, where the particle fields are in a symmetric
state at high temperature and a broken symmetry state at low temperature
\cite{Kirzhnits1972,PhysRevD.9.3320,PhysRevD.9.3357}. With regard to
stabilization against the parametric instability, our theoretical analysis and
numerical simulations show that thermal dissipation can stabilize the false
vacuum state against the instability, allowing one to observe decay of the false
vacuum state via a thermal first-order phase transition. However, for realistic
experimental parameters the damping rate required for stabilization is
significantly in excess of the rate predicted within the SPGPE description close
to equilibrium. We discuss some of the possible caveats around taking this rate
too literally in our non-equilibrium experimental scenario. We conclude that
experiments measuring damping rates in non-equilibrium, quasi-1D Bose gases,
would be helpful to settle the question definitively.

The remainder of the paper is structured as follows. The system is described in
section \ref{system}. A theoretical analysis in section \ref{theory} explores
the parametric resonance phenomenon and demonstrates that thermal damping is
able to suppress the resonance.  A numerical investigation, including the
initialisation stage and showing bubble nucleation, is presented in section
\ref{numerical}. Section \ref{experimental} gives physical values for the
parameters based on a particular atomic species, and improvements to the
present scenario are described in the concluding section.

\section{System}
\label{system}

 \begin{center}
\begin{figure}[htb]
  \includegraphics[width=0.4\columnwidth]{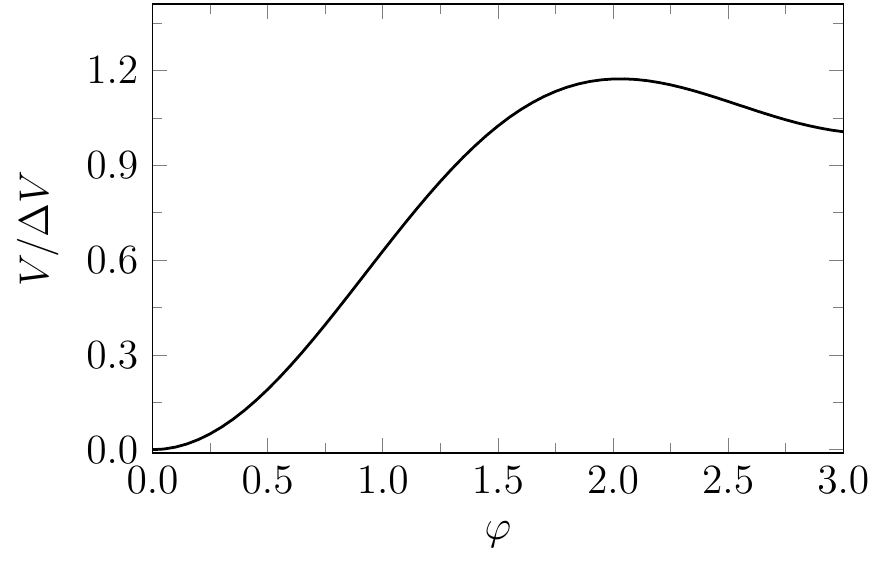}
\caption{The field potential $V$ plotted as a function of the
relative phase of the two atomic wave functions, $\varphi$. The
metastable phase is at the minimum $\varphi=\pi$ and the stable
phase is at the global minimum $\varphi=0$. The difference in energy
density between these phases is $\Delta V$.} \label{pot} 
\end{figure} 
\end{center} 

The system is a one-dimensional, two-component Bose gas of atoms with mass $m$.
The two components are different spin states of the same species, separated in
energy by an external magnetic field and coupled by a time-modulated microwave field. 
The Hamiltonian is given by
\begin{equation}
H=\int d^nx\left\{
-{\hbar^2\over 2m}\psi^\dagger\nabla^2\psi
+V_{\rm osc}(\psi,\psi^\dagger)\right\}.
\label{hamil}
\end{equation}
where the field operator  $\psi$ has two components $\psi_i$, $i=1,2$. The trapping
potential has been omitted, as we are interested in geometries where this is uniform
across the system. The potential $V_{\rm osc}$ represents the atomic interactions,
\begin{equation}
V_{\rm osc}=\frac{g}2\sum_i\left(\psi_i^\dagger\psi_i\right)^2 - \mu \psi^\dagger \psi
-(\mu\epsilon^2+\delta\hbar\omega\cos\omega t)\,\psi^\dagger\sigma_x \psi,
\label{potential}
\end{equation}
where $\sigma_{\{x,y\}}$ are Pauli matrices.
The interaction potential includes the chemical potential $\mu$, 
and equal intra-component $s$-wave interactions of strength $g$ between the field operators. 
(We assume inter-component $s$-wave interactions have been suppressed by tuning 
the external magnetic field.)
Microwave-induced interaction terms have a constant contribution determined by a
(small) dimensionless parameter $\epsilon$ and
a modulation depending on another (small) parameter $\delta$. When the microwave interaction 
terms are switched off, the potential has a degenerate vacuum state with  
$\langle \psi_1^\dagger\psi_1\rangle =\langle\psi_2^\dagger\psi_2\rangle =\rho_0$,
where the mean number density $\rho_0=\mu/g$. 
The microwave interaction terms break the degeneracy of the
lowest energy state, resulting in a global minimum of the potential 
with $\psi_1=\psi_2$, and a saddle point at  $\psi_1=-\psi_2$.

By applying a quantum mechanical averaging procedure \cite{FialkoFate2015,FialkoUniverse2017}, 
Fialko et al. arrived at an effective theory that is a candidate for describing the system on timescales longer 
than $2\pi/\omega$. The effective potential $V_{\rm static}$, that replaces $V_{\rm osc}$, has an extra quartic interaction,  
\begin{equation}
V_{\rm static}=\frac{g}2\sum_i\left(\psi_i^\dagger\psi_i\right)^2 - \mu \psi^\dagger \psi
-\mu\epsilon^2\psi^\dagger\sigma_x \psi+
\frac{g}{4}\lambda\epsilon^2(\psi^\dagger\sigma_y\psi)^2,\label{vstat}
\end{equation}
where $\lambda=\sqrt{2}\delta/\epsilon$. The modulation has done its job in 
converting the stationary point at $\psi_1=-\psi_2$ into a local minimum in the potential when 
$\lambda>1$. 
We can illustrate this by introducing the relative 
phase $\varphi$ between the spin components, 
such that the mean field $\psi_1\approx\rho_0 e^{i\varphi/2}$ and $\psi_2\approx\rho_0 e^{-i\varphi/2}$, 
then the potential becomes
\begin{equation}
V_{\rm static}\approx-2\mu\epsilon^2-2\mu\epsilon^2\cos\varphi+\mu\epsilon^2\lambda^2\sin^2\varphi.
\end{equation}
The potential has a true vacuum state at $\varphi=0$ and a false vacuum
at $\varphi=\pi$, as shown in Fig. \ref{pot}. The height of the potential barrier between the
minima is determined by $\lambda$ and $\epsilon$, and fluctuations in the phase
of the wave functions are far larger than fluctuations in the modulus when $\epsilon$ is small.

In our version of the experimental proposal, the system is initially prepared in the metastable
phase at a temperature $T$. Two important parameters
for the system are the healing length $\xi=\hbar/(mg\rho_0)^{1/2}$
and the sound speed $c=\hbar/(m\xi)$. In one dimension, the physics of Bose gases
critically depends also on the dimensionless interaction strength parameter, $\zeta =
(\rho_0 \xi)^{-2}$, and the temperature 
\cite{KheruntsyanPair2003,KheruntsyanFinite2005, BouchouleTwoBody2012, HenkelCrossover2017}. 
We consider the weakly interacting case, $\zeta \ll 1$. A phase-fluctuating
quasi-condensate, in which density fluctuations are suppressed, appears at
temperatures below the cross-over temperature \footnote{Note that our definition
  omits a numerical factor 2 often found elsewhere~\cite{KheruntsyanPair2003,
    KheruntsyanFinite2005, BouchouleTwoBody2012}.}
\begin{equation}
T_{CO}={\hbar c\rho_0\over k_B}.\label{TCO}
\end{equation}
The gas remains degenerate up to a temperature of order
$T_D=\zeta^{-1/2}T_{CO}>T_{CO}$. We shall show that the atomic gas will behave
as a relativistic Klein-Gordon system undergoing a first order phase transition at a temperature
around a few percent of $T_{CO}$.

\section{Theoretical discussion}\label{theory}

We will model the spinor gas with the SPGPE for the stochastic two-component field 
$\psi$ and its conjugate $\overline\psi$. From this point in the paper, we will use the healing length
$\xi$ to define the length unit, and the characteristic frequency 
$\omega_0=c/\xi$ to define the time unit. The potential is measured in units of 
$\hbar\omega_0\rho_0$. In this section, our considerations are independent of
the projection involved in the SPGPE, so for simplicity we remove it and consider a
stochastic GPE (SGPE), which in dimensionless form is
\begin{equation}
  i\dot\psi=(1-i\gamma) \left(-\frac12\nabla^2\psi+{\partial V\over\partial \overline \psi}\right)+\eta,
\label{spgpe}
\end{equation}
where $V$ is chosen for either the oscillating potential or the averaged
potential. An SGPE of this general form is another well-established stochastic
description of a finite-temperature Bose gas \cite{StoofCoherent1999,
  StoofDynamics2001}. In the oscillating case, 
\begin{equation}
V_{\rm osc}=\frac12\sum_i\left(\overline\psi_i\psi_i-1\right)^2-
(\epsilon^2+\delta\omega\cos\omega t)\,\overline\psi\sigma_x\psi.
\end{equation}
The stochastic noise term $\eta$ has a Gaussian distribution, with variance
\begin{equation}
\langle\eta(x,t)\overline\eta(x',t')\rangle=2\gamma T\delta(x-x')\delta(t-t').
\end{equation}
In one dimension, the temperature is measured in units of the cross-over
temperature $T_{CO}$ defined in Eq. (\ref{TCO}). (In $n$ dimensions,
the unit of temperature is $\hbar\omega_0\zeta^{-1/2}/k_B$, where $\zeta=(\rho_0\xi^n)^{-2}$
is the coupling strength.)

Parametric instability sets in for wavelengths
shorter than the healing length and lying in narrow resonance bands. In this section,
we will verify this assertion by linearising the SGPE and using averaging
techniques. We will examine the role that friction plays in damping out the
parametric resonance and restoring the behavior seen for a static
potential. The fully non-linear system will be analysed numerically and compared to
the time averaged system in the next section.

\subsection{Simple averaging}

A simple averaging procedure can be used to produce effective equations describing the
system on time scales large compared to the modulation timescale. 
Previous discussions of time averaging in the modulated system 
\cite{FialkoFate2015,FialkoUniverse2017} have shown that the system
can be described by the effective Hamiltonian with the static potential (\ref{vstat}),
which can be further reduced to a Klein-Gordon theory. Starting from the effective Hamiltonian,
in Ref. \cite{Billam:2020xna}, it was shown that the SPGPE with the static
potential reduces to a stochastic, damped Klein-Gordon theory. In order to complete this picture, we
show below how the SPGPE with the oscillating potential reduces to the
damped Klein-Gordon theory.

A re-parameterisation of the wave functions can be used to facilitate the linearisation,
\begin{align}
\psi_1&=+e^{\chi/2}e^{\sigma/2}e^{i\varphi/2}e^{i\theta/2},\\
\psi_2&=\pm e^{\chi/2}e^{-\sigma/2}e^{-i\varphi/2}e^{i\theta/2}.
\end{align}
The positive and negative signs are chosen for expansion about the true or false vacuum
respectively. We assume $\gamma=O(\epsilon)$ and $\delta=O(\epsilon)$, where $\epsilon$
is the RF mixing parameter introduced in the potential (\ref{potential}).
At leading order, the SPGPE given in Eq. (\ref{spgpe}) reduces to linear 
(Bogoliubov-de Gennes) equations for the four fields 
$\chi$, $\sigma$, $\varphi$ and $\theta$.
The spatial Fourier transform of the relative phase $\varphi$ 
couples only to the relative density variation $\sigma$,
\begin{align}
\dot\varphi&=-(a\pm2\delta\omega\cos\omega t)\sigma-
\gamma (b\pm2\delta\omega\cos\omega t)\varphi+\eta_\varphi,\label{eq1}\\
\dot\sigma&=(b\pm2\delta\omega\cos\omega t)\varphi-
\gamma (a\pm2\delta\omega\cos\omega t)\sigma+\eta_\sigma.\label{eq2}
\end{align}
The noise terms $\eta_\varphi$ and $\eta_\sigma$ are independent Gaussian 
random fields with variance $2\gamma T$.
The coefficients $a$ and $b$ are functions of the wave number $k$,
\begin{align}
a&=\frac12(k^2+4\pm4\epsilon^2),\label{aeq}\\
b&=\frac12(k^2\pm 4\epsilon^2)\label{beq}.
\end{align}
Free oscillations of the un-modulated system have frequency $2\sqrt{ab}$,
leading to instability of  the false vacuum for small $k$.

Eqs. (\ref{eq1}) and (\ref{eq2}) are analogous to those which describe the stabilisation 
of the inverted Kapitza pendulum \cite{FialkoFate2015,FialkoUniverse2017}, and we follow a 
similar line of analysis to find a time-averaged effective theory. We set
\begin{align}
\varphi&=\varphi_0+\varphi_1\cos\omega t+\varphi_2\sin\omega t,\\
\sigma&=\sigma_0+\sigma_1\sin\omega t+\sigma_2\cos\omega t,
\end{align}
where we assume $\omega^2\gg ab$ and the functions $\varphi_n$, $\sigma_n$ are slowly varying
compared to the oscillatory terms. 
Taking the coefficients of the $\sin\omega t$ and $\cos\omega t$ terms in Eqs. (\ref{eq1}) and (\ref{eq2}) gives
\begin{align}
\varphi_1&={a\over\omega}\sigma_1={2\delta a\over\omega}(\varphi_0-\gamma\sigma_0),\\
\sigma_2&=-{b\over \omega}\varphi_2={2\delta b\over\omega}(\sigma_0+\gamma\varphi_0).
\end{align}
After substituting these back into Eqs. (\ref{eq1}) and (\ref{eq2}), and taking the period 
averages, we arrive at equations for the slowly varying terms $\varphi_0$ and $\sigma_0$.
Dropping $O(\gamma\delta^2)$ terms, and taking the long-wavelength limit $k \ll 1$, the leading terms are
\begin{align}
\dot\varphi_0&=-2\sigma_0+\overline\eta_\varphi,\label{eq3}\\
\dot\sigma_0&= \frac12\omega_k^2\varphi_0-2\gamma\sigma_0+\overline\eta_\sigma,\label{eq4}
\end{align}
where $\omega_k^2=k^2+m_\varphi^2$ and the mass $m_\varphi=2\epsilon(\lambda^2\pm 1)^{1/2}$.
By eliminating $\sigma_0$, we find that the system is equivalent to a 
damped Klein-Gordon field with noise $\eta_{\rm eff}$,
\begin{equation}
\ddot\varphi_0+2\gamma\dot\varphi_0+\omega_k^2\varphi_0=
\eta_{\rm eff}.
\label{dkg}
\end{equation}
The false vacuum is stable against small fluctuations provided that the mass is real, i.e. $\lambda>1$.
In Ref. \cite{Billam:2020xna}, the damped Klein-Gordon system
of Eqs. (\ref{eq3}) and (\ref{eq4}) was obtained starting from the 
effective Hamiltonian with a static potential, $V_{\rm static}$, and then linearising the SGPE.

If the condition that $\omega^2\gg ab$ is dropped, then secular terms can arise which
invalidate the averaging procedure used above. In order to examine this possibility, we
turn to the regime of parametric resonance.

\subsection{Parametric resonances}

\begin{center}
\begin{figure}[htb]
\begin{center}
\scalebox{0.45}{\includegraphics{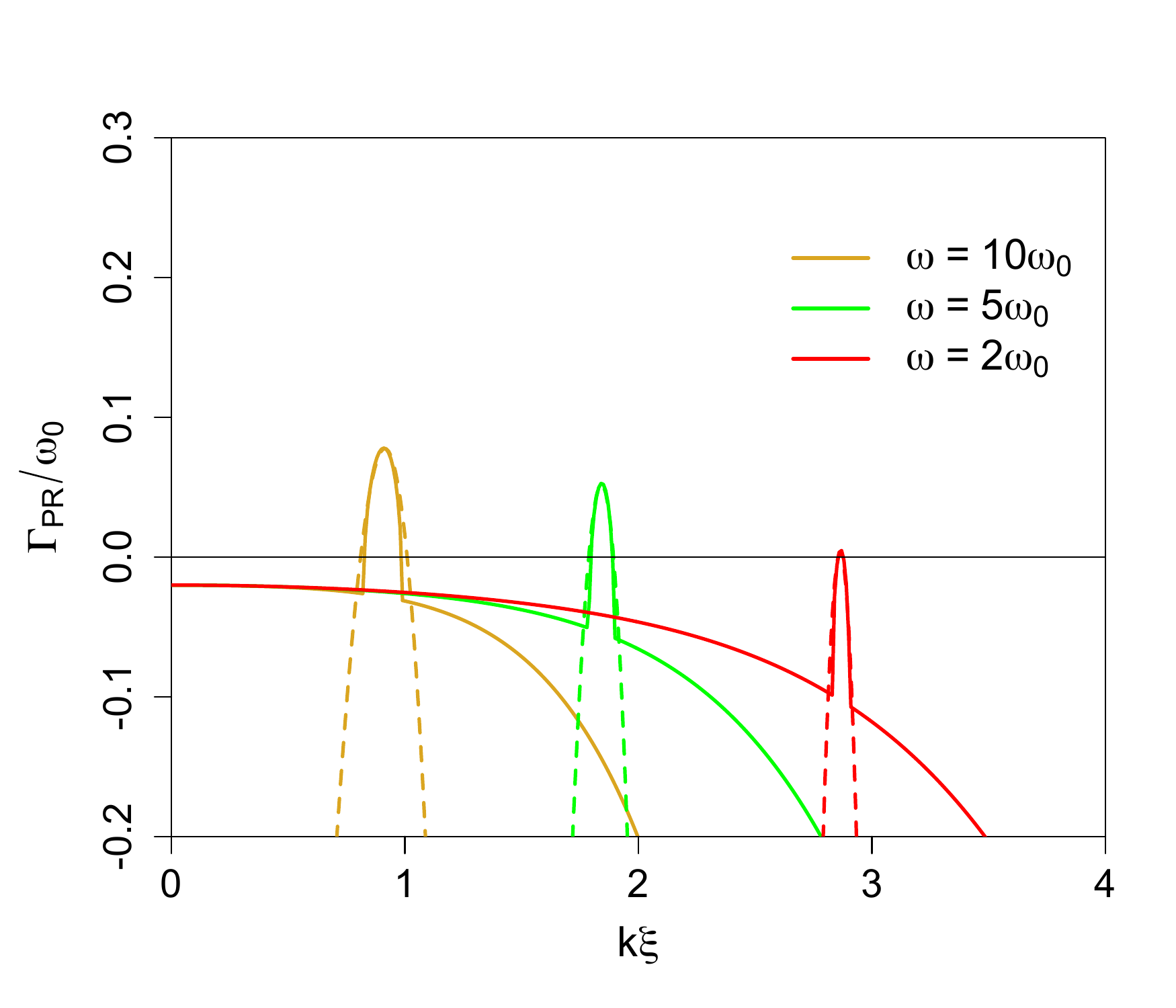}}
\scalebox{0.45}{\includegraphics{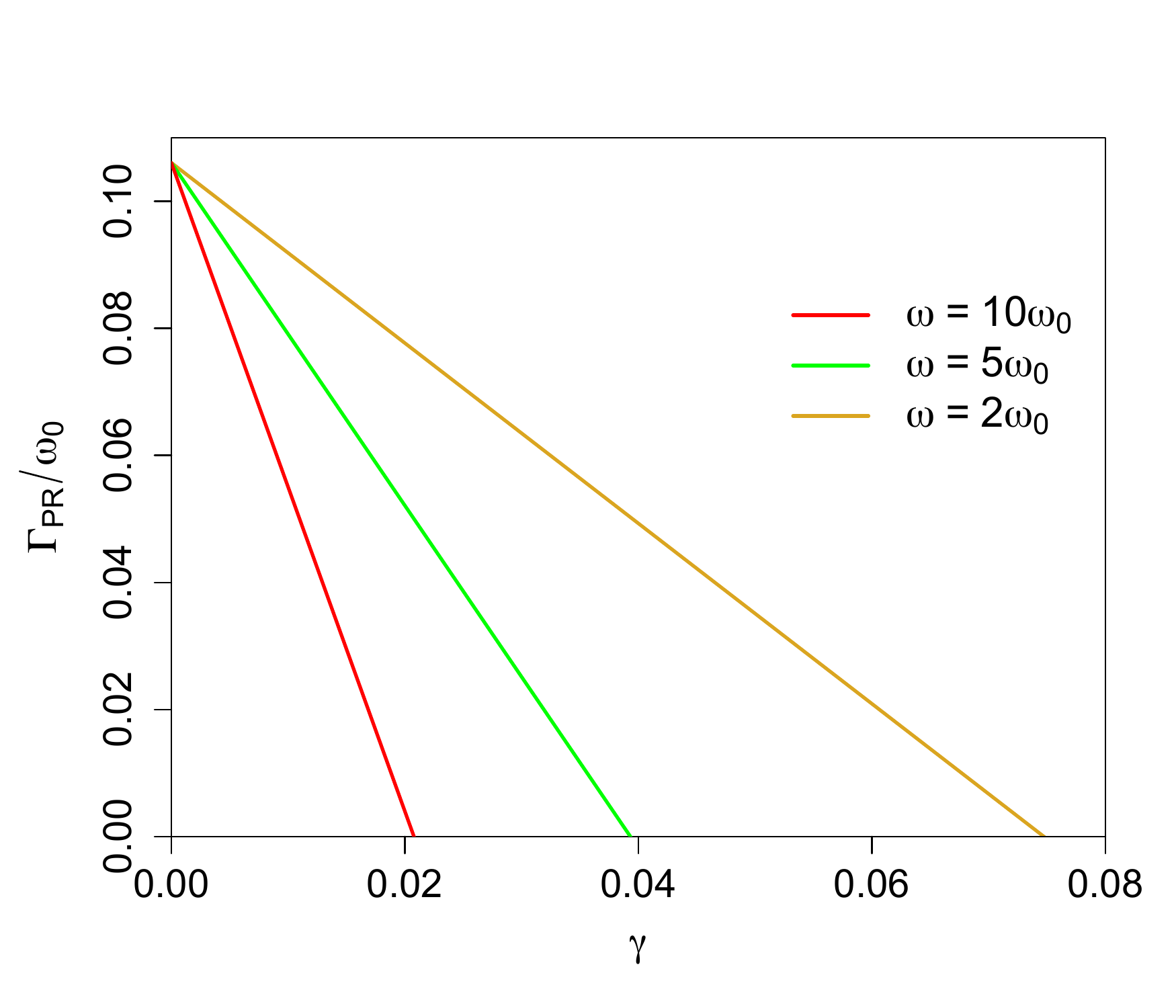}}
\end{center}
\caption{On the left, the growth rate $\Gamma_{PR}$ of the spatial Fourier modes,
plotted as a function of wave number $k$. The microwave mixing is modulated with
frequency $\omega$ and the damping is governed by the parameter $\gamma$.
In this figure, $\gamma=0.02$. The resonance band approximation 
Eq. (\ref{Gamma}) is shown by a dotted line. 
On the right, the growth rate at the top of the resonant peak plotted as a function of the
damping $\gamma$. In each case, the amplitude of the modulation corresponds 
to $\lambda=1.5$.
}
\label{damping}
\end{figure}
\end{center}

The system with the oscillatory potential is known to feature a parametric resonance
\cite{Braden:2017add,Braden:2019vsw} which destabilises the false vacuum when $k\sim 1$ in
healing length units.
The analysis in the earlier work did not take into account the effects
of damping, which may act to reduce, or even remove, the parametric resonance
allowing first order decay phenomena such as bubble nucleation to show up.
We shall now consider parametric resonance in the unforced system with damping.
The forced system with noise and parametric amplification is covered in the appendix.

Close to the first resonance, 
which has frequency $\omega/2$, we approximate the relative phase by
\begin{align}
\varphi&=A(t)\cos \left(\omega t/2 \right) +B(t)\sin \left(\omega t/2 \right),\label{resphi}\\
\sigma&=C(t)\cos \left( \omega t/2 \right) +D(t)\sin \left( \omega t/2 \right),
\end{align}
where $A\dots D$ are slowly varying functions of time.
The time averages of (\ref{eq1}) and (\ref{eq2}) give
equations for $A\dots D$,
\begin{align}
\dot A&=-\gamma b A-\frac\omega2 B-(a+\delta\omega)C,\label{eqA}\\
\dot B&=\frac\omega2 A-\gamma b B+(\delta\omega-a)D,\label{eqB}\\
\dot C&=(\delta\omega+b)A-\gamma a C-\frac\omega2 D,\label{eqC}\\
\dot D&=-(\delta\omega-b)B+\frac\omega2 C-\gamma a D.\label{eqD}
\end{align}
The growth rate $\Gamma_{PR}$ of the solutions is determined by the eigenvalue of
this linear system with the largest real part, $\lambda_+$.
This depends on the wave number $k$ included in $a$ and $b$. 
The solutions only grow in a narrow band of $k$ values around $k=k_b$, where
the forcing and the natural frequency coincide, i.e. $\omega^2=4ab$.
Inserting the $a$ and $b$ from Eqs. (\ref{aeq}) and (\ref{beq}), the centre of the 
resonance band is at
\begin{equation}
k_b^2=2(1+\omega^2/4)^{1/2}-2\mp4\epsilon^2.\label{kb}
\end{equation}
Order $\epsilon^2$ terms are small and we can discard them.
Inside the resonance band, we let $k=k_b+\Delta k$, where $\Delta k$ is small.
The eigenvalues of the system are parabolic in $\Delta k$,
\begin{equation}
\Gamma_{PR}=2\delta-\frac12\gamma(k_b^2+2)-{(k_b^2+2)^2\over 4\delta(k_b^2+4)}\Delta k^2.\label{Gamma}
\end{equation}
The growth rate in the centre of the band where $\Delta k=0$ is therefore $2\delta-(k_b^2+2)\gamma/2$.
Using Eq. (\ref{kb}) for $k_b$, we see that the resonance is damped out when
\begin{equation}
\gamma>{4\delta\over \sqrt{4+\omega^2}}.\label{gammalim}
\end{equation}
Since $\delta=\lambda\epsilon/\sqrt{2}$, the resonance is damped out for values
of the friction $\gamma$ of order $\epsilon$, which we have taken to be small.

Figure \ref{damping} shows the growth rate of the modes with different values of the damping
and modulation frequency, obtained directly from the eigenvalues of Eqs. (\ref{eqA}) and (\ref{eqB})
with no approximations. The underlying damping effect agrees with the damped Klein-Gordon
system in Eq. (\ref{dkg}). The resonance bands are in good agreement with the approximation (\ref{Gamma}). 
Higher order resonance bands have slower growth rates, and the damping is enhanced by larger $k$ values.
Consequently, fluctuations in the higher resonance bands are damped out if the fluctuations in 
the first resonance band are damped.

\section{Numerical Results}\label{numerical}

We now perform numerical simulations of the SPGPE to examine how the decay of
the false vacuum proceeds in the fully non-linear system. We compare the
dynamical phase transition for both the oscillating and static potential, for
parameter ranges where we expect to see parametric instability, hoping to see a
progression towards the first order behaviour that occurs in the static case.

In this section we use the simple growth SPGPE \cite{GardinerStochastic2002,
GardinerStochastic2003, bradley_bose-einstein_2008, blakie_dynamics_2008} as
extended to multi-component and spinor condensates in
Ref.~\cite{BradleyStochastic2014}. Including the projector, the dimensionless
SPGPE reads 
\begin{equation}
  i \, d\psi= \mathcal{P} \left\{ (1-i\gamma) 
  \left(-\frac12\nabla^2\psi+{\partial V\over\partial \overline \psi}\right) \, dt + d\eta \right\},
  \label{spgpe_with_p}
\end{equation}
where the noise is correlated at equal times
as $\langle d\eta(x,t)d \overline \eta(x',t)\rangle=2\gamma T\delta(x-x') dt$.  The projection
${\cal P}$ in the
SPGPE cuts off modes with wave number $k>k_c$, where $\hbar k_c=(2m
k_BT)^{1/2}$ in physical units; this removes sparsely-populated high-momentum
modes, not well-described by the classical field approximation, from the
simulation~\cite{blakie_dynamics_2008}.  We take a one dimensional system with
periodic boundary conditions, such as would be seen in a cold atom ring trap.

\begin{center}
\begin{figure}[htb]
\begin{center}
 \includegraphics[width=0.4\columnwidth]{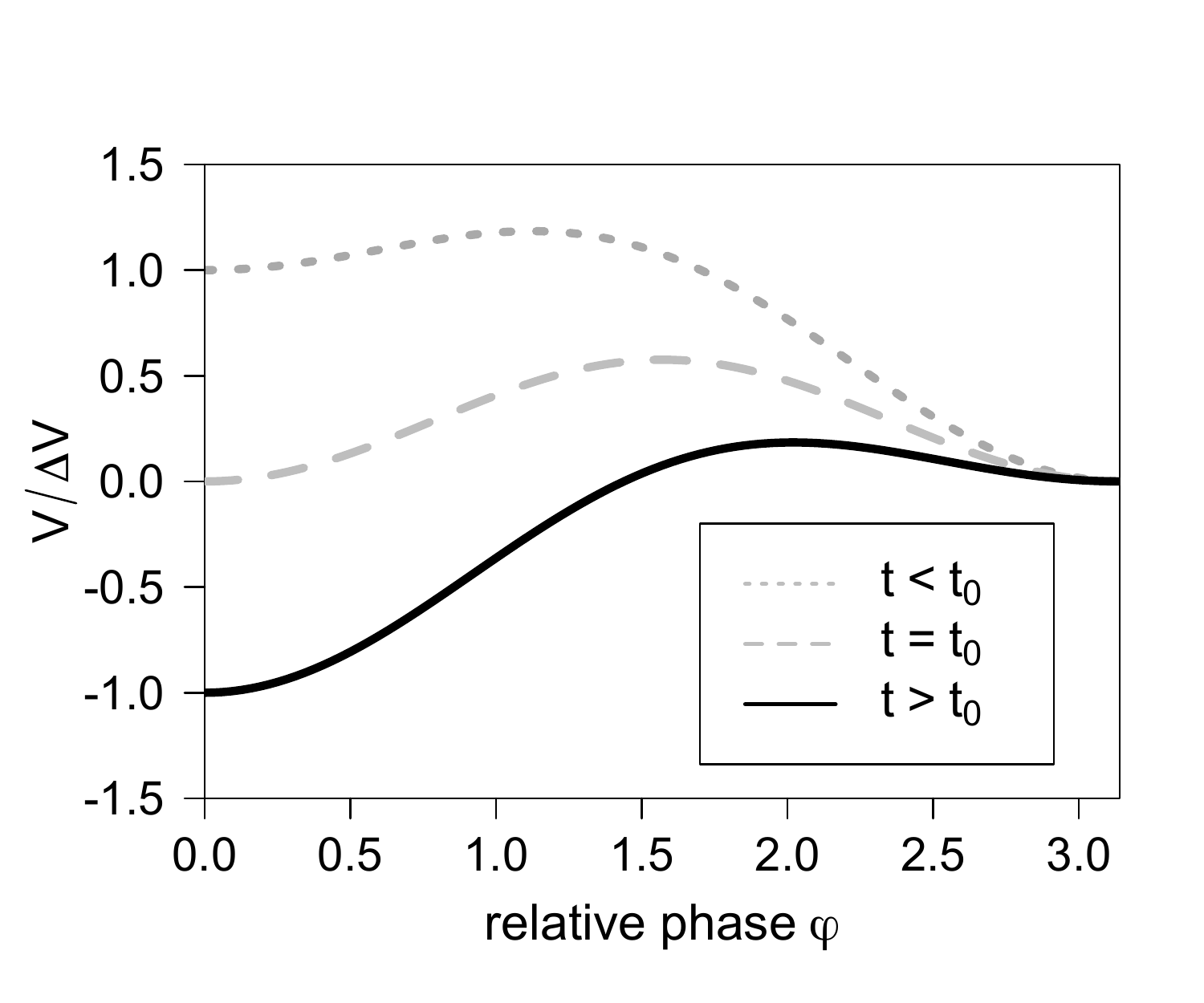}
\end{center}
\caption{The potential switches from having a global minimum at $\varphi=\pi$ for $t<t_0$
to a local minimum when $t>t_0$.
}
\label{switch}
\end{figure}
\end{center}

An experimental protocol is required to initialise the system in a metastable
phase. For the numerical simulations, we have implemented a procedure in which
a minimum of the potential switches from a global to a local minimum as a
control parameter, $\alpha(t)$, is changed. Conceptually, the system is loaded
into the stable phase in the initial global minimum, with thermal fluctuations
at temperature $T$. After a settling down period, the potential is adjusted
using the control parameter so that the system enters the metastable phase. 

In more detail, this initialisation step is achieved by including $\alpha(t)$
in the RF mixing terms in the interaction potential:
\begin{equation}
V_{\rm switch}=\frac12\sum_i\left(\overline\psi_i\psi_i-1\right)^2-
\left[\epsilon^2\cos\alpha(t)+\delta\omega\cos\omega t\right]\,\overline\psi\sigma_x\psi.
\end{equation}
The RF mixing has both an oscillatory contribution with frequency $\omega$ and a
slowly varying contribution with the control parameter $\alpha(t)$. Both terms are
determined by the voltage input into the RF antenna.  We set up $\alpha(t)$ so
that
\begin{equation}
\alpha=\frac\pi2+\frac\pi2{\rm tanh}\left({t-t_0\over \tau}\right).
\end{equation}
This ensures that the state with relative phase $\varphi=\pi$ is a stable
minimum of the potential for $t<t_0$ and a metastable minimum for $t>t_0$, as
shown in figure \ref{switch}.  The switching time $\tau$ is chosen to be larger
than the timescale $1/\omega$, in order not to interfere with the time
averaging, but chosen shorter than the bubble nucleation timescale so that the
results reflect the properties of decay in the final potential rather than any
transient effects.

\begin{center}
\begin{figure}[htb]
\begin{center}
 \includegraphics[width=0.8\columnwidth]{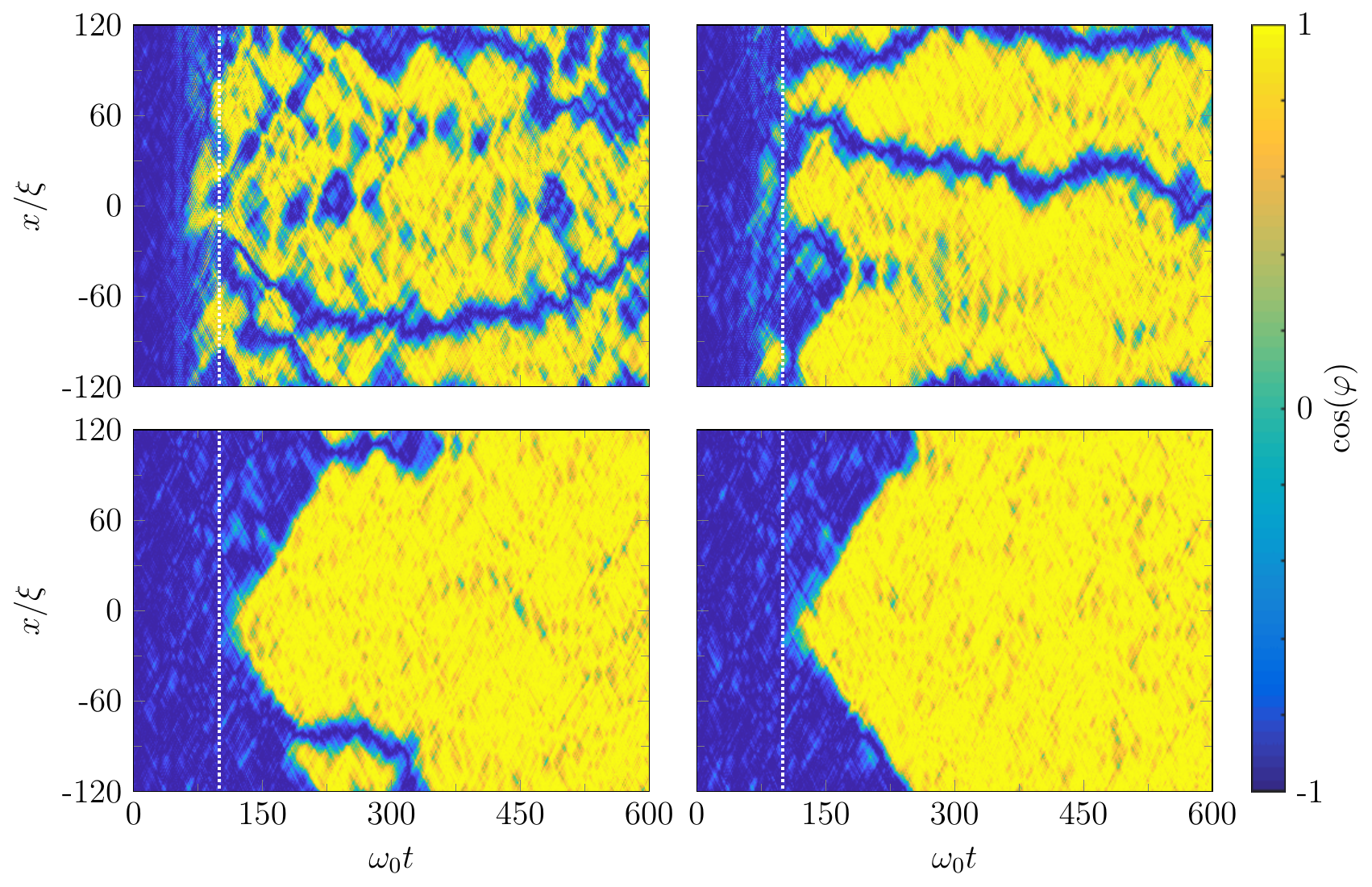}
\end{center}
\caption{This series of images shows typical phase transition scenarios
with different values of the damping $\gamma$ in the oscillating system. 
The system is prepared in
a thermal state with $\varphi=\pi$ which is converted to a metastable
state at $\omega_0t=100$ (indicated by white vertical line), with switching time $\omega_0 \tau = 5$.
The top row, with $\gamma=0.005$ (left) and $\gamma=0.01$ (right),
shows no evidence of bubble nucleation.
The second row, with $\gamma=0.02$ (left) and $\gamma=0.03$ (right), 
clearly shows bubbles nucleating. By the final image, the nucleation events are 
indistinguishable from nucleation with a static 
potential. In these runs, the microwave mixing $\epsilon=0.05$, and the modulation 
has amplitude $\lambda=1.5$ and frequency $\omega=5\omega_0$.
}
\label{images}
\end{figure}
\end{center}

After the potential is switched at $t=t_0$, our simulations show that the
system nucleates false vacuum regions. Examples are shown in Fig.
{\ref{images}. A qualitative dependence on the damping $\gamma$ appears for the
oscillating potential in the parametric resonance regime. In the low damping
case, there is hardly any sign of bubble nucleation; rather, the relative phase displays
strong fluctuations after a rapid growth of the instability.  As the the damping is
increased, the onset of bubble nucleation is a clear indication of first order
behaviour. Above a certain value of $\gamma$, there is no longer any apparent
difference between the oscillating potential and the static potential.

\begin{center}
\begin{figure}[htb]
\begin{center}
 \includegraphics[width=0.4\columnwidth]{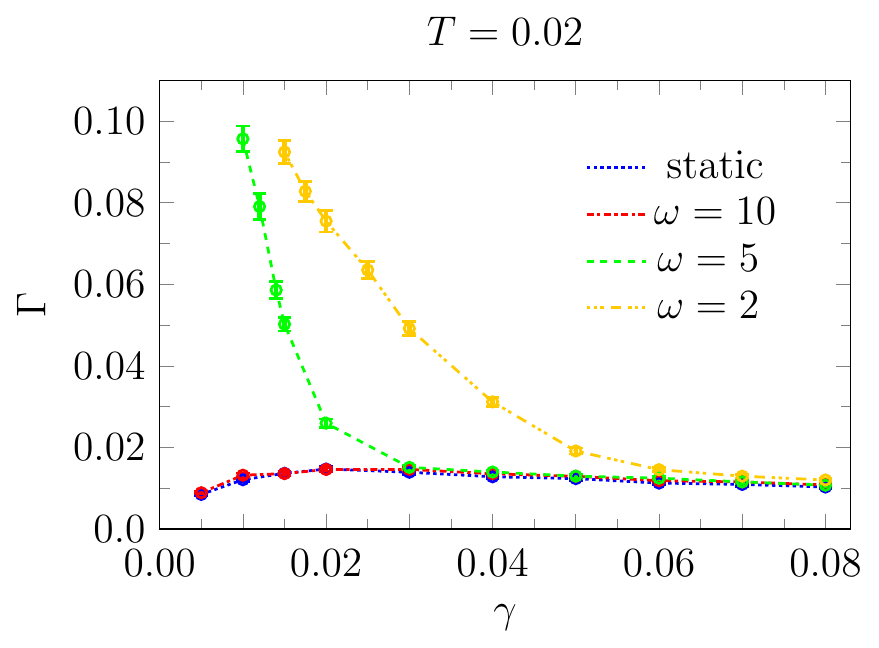}
 \includegraphics[width=0.4\columnwidth]{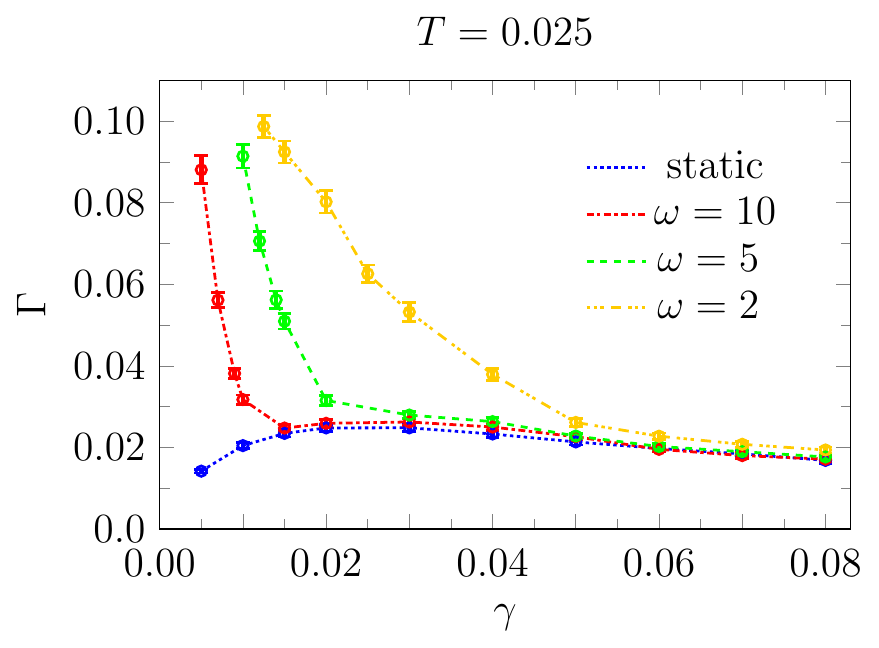}\\
  \includegraphics[width=0.4\columnwidth]{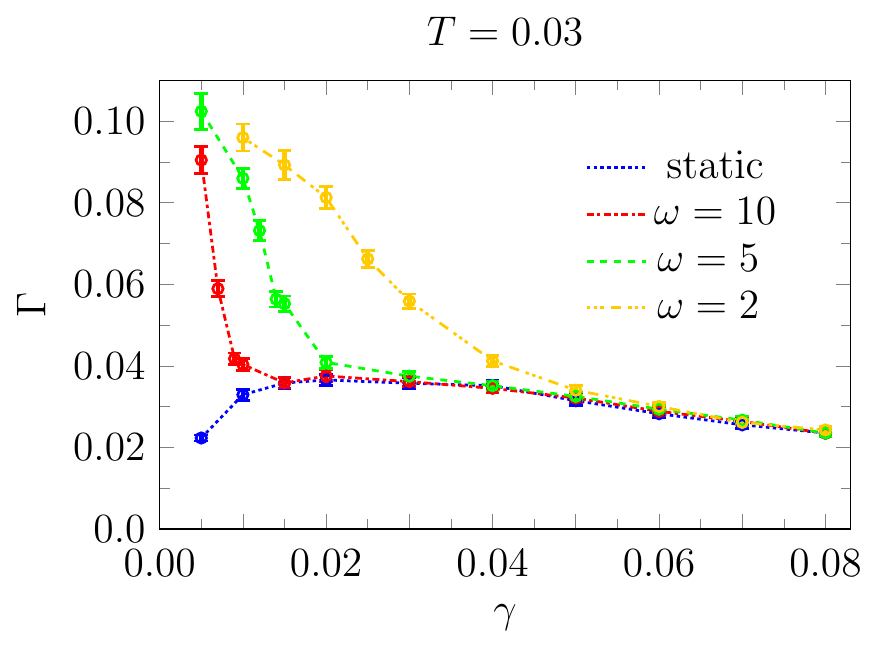}
\end{center}
\caption{The decay rate is shown as a function of the friction $\gamma$ for the
oscillating potential and the static potential. The rates for the oscillatory
potential are larger at small $\gamma$, but agree with the rates for the static
potential at larger $\gamma$. In these plots the microwave mixing is
$\epsilon=0.05$ and the modulation $\lambda=1.5$.  The temperatures are
$T=0.02T_{CO}$, $T=0.025T_{CO}$ and $T=0.03T_{CO}$.
}
\label{rate}
\end{figure}
\end{center}

In order to make a quantitative comparison between the oscillating potential
and the static potential, we compare the decay rate of the metastable state in
the two cases. The time it takes for the phase transition is measured by
observing the spatial average $\langle\cos\varphi\rangle$ to be larger than
$-1+\Delta$, where $\Delta=0.9$ is chosen to be much larger than the typical
fluctuations of $\langle\cos\varphi\rangle$ due to thermal noise in the system.
Running many stochastic trajectories allows us to compute the probability, $P$,
of remaining in the metastable state at time $t$. A fit to the exponential form $P=a
e^{-\Gamma t}$ over the time intervals seen to be exhibiting exponential decay (we find this to be times late enough that $P < 0.7$) yields the decay rate $\Gamma$. Error bars are estimated using a bootstrap procedure 
as described in~\cite{BillamSimulating2019}. The decay rate is plotted in
Fig. \ref{rate} both for the oscillating potential, with a range of modulation
frequencies, and for the static potential.

As expected from the theoretical discussion, there is a convergence in the
decay rates around a value of the friction given in Eq. (\ref{gammalim}). 
The first order behaviour sets in when the parametric resonance is
less significant than the bubble nucleation, which occurs around 
$\Gamma\approx \Gamma_{PR}$. There is good agreement
with the theoretical predictions shown in Fig. \ref{damping}.

In one of the examples, $T=0.02 T_{CO}$ and $\omega=10\omega_0$,
there is no apparent difference between the
oscillating potential and the static potential. In this example, the removal
of modes with $k>k_c$ in the SPGPE has eliminated the resonance band. This case
is similar to some of the examples in Ref. \cite{Braden:2019vsw}, where there was
an effective cutoff from the size of the spatial grid used in the numerical
simulations. We therefore have agreement between the two sets
of results for this case. 

In cases close to the point where the resonance is suppressed by the
cutoff we do not expect our model to be quantitatively correct, since small
changes in the cutoff would have a major effect on the result. Such cutoff-dependent effects do not correspond to
a physical Bose gas where there is no hard cutoff. However, this does raise the interesting
question of what happens if the resonance is significantly above the cutoff
wavenumber. In this regime one might expect the resonance to affect the thermal
cloud more than it affects the quasi-condensate directly. Thermal cloud
dynamics are outside the scope of our present SPGPE study. However,
investigation of whether first-order behaviour can be seen in this regime for
low damping is an interesting avenue for further work using other models that include thermal cloud dynamics~\cite{ProukakisFiniteTemperature2008}.

\section{Experimental realisation}\label{experimental}

We assume a cigar-shaped Bose gas tightly confined by a transverse harmonic
trap of frequency $\omega_\perp$. Assuming a three-dimensional thermal cloud,
the one-dimensional SPGPE above \eqnrefp{spgpe_with_p} with
dimensionally-reduced interaction strength $g = 2\hbar a_s\omega_\perp$ is a
valid description provided $\hbar \omega_\perp \lesssim k_\mathrm{B} T$ \cite{BradleyLowDimensional2015}. In
principle, one should have chemical potential $\mu=\hbar\omega_0 \ll \hbar
\omega_\perp$. However, in practice $\omega_0 \lesssim\omega_\perp$ has been
found sufficient in 1D SPGPE equilibrium
studies~\cite{CockburnQuantitative2011, DavisYangYang2012} of quasi-1D
atom-chip experiments~\cite{TrebbiaExperimental2006, vanAmerongenYangYang2008}.

As an example experimental configuration, we consider one of the experimental
setups proposed by Fialko et al. \cite{FialkoUniverse2017}, which is based on
tuning the interactions between two Zeeman states of ${}^{41}{\rm K}$. 
The interactions can be tuned using a Feshbach
resonance to achieve the required close-to-zero inter-component scattering
length~\cite{FialkoUniverse2017}. Based on the intra-component scattering
length $a_s=60$ Bohr radii, suitable experimental parameters would be $2.4\times10^4$ atoms in a
quasi-1D optical trap \cite{SalcesCarcobaEquation2018} of length $\SI{218}{\micro
m}$ and transverse frequency $\omega_\perp=2\pi \times \SI{428}{Hz}$. 
The frequency $\omega_0$ in the suggested configuration has a value 
around $2\pi \times \SI{300}{Hz}$, and satisfies the above constraint.
The interaction strength $\zeta=10^{-4}$, and the cross-over
temperature $T_{CO}= \SI{44}{\micro K}$. In this context the results in
\figreft{rate}{} correspond to temperatures of $\SI{28.8}{\nano K}$, $\SI{36}{\nano K}$
and $\SI{43.1}{\nano K}$, where bubble nucleation should be observable.
For these parameters, the bubbles in Figs. \ref{images} and \ref{rate} are nucleating around $25$ 
milliseconds or so after the potential is made metastable.

Another realisation could be achieved with two Zeeman states of ${}^{7}{\rm Li}$
\cite{FialkoUniverse2017}.
In this case the intra-component scattering lengths of the two states are different, so
that the analysis would need to be generalised. However, using the mean scattering
length $a_s=10$ Bohr radii as a guide, a gas of $4.8\times 10^{4}$ atoms in a trap of length 
$\SI{102}{\micro m}$ with transverse frequency $\omega_\perp=2\pi \times \SI{16.1}{\kilo Hz}$
has cross-over temperature $T_{CO}=\SI{77.3}{\micro K}$, interaction
parameter $\zeta=2.5\times 10^{-5}$, and frequency scale
$\omega_0=2\pi \times \SI{8.06}{\kilo Hz}$.

While we have shown in this work that thermal damping can suppress the unwanted
resonance effects in principle, the experimental value of $\gamma$ is crucial to
whether or not they are actually suppressed in a given experiment. In
equilibrium $\gamma$ can be predicted \textit{a priori} within the SPGPE
theory~\cite{bradley_bose-einstein_2008, blakie_dynamics_2008}, and
near-equilibrium experiments have been quantitatively descibed using this
\textit{a priori} value of $\gamma$ \cite{rooney_persistent-current_2013}.
However, further from equilibrium SPGPE studies have typically achieved a
better match to experiments by treating $\gamma$ as a free parameter that may
vary significantly from the \textit{a priori} value. Effective $\gamma$ values up to $\gamma
= 0.02$ have been employed to match experiments in this way
\cite{weiler_spontaneous_2008,ota_collisionless_2018,liu_dynamical_2018}.

In the experiments proposed here, bubbles nucleate from a thermal equilibrium
state that has been raised to metastability by our control parameter protocol.
This would appear to be a reasonably near-equilibrium scenario up until the
point a bubble is nucleated and begins to grow. With the above example experimental 
parameters for ${}^{41}{\rm K}$, the {\it a priori} prediction for our two-component 
SPGPE model is $2.5\times 10^{-5}\leq\gamma\leq 4.7\times 10^{-5}$.
This stated range includes the temperature difference from $0.02T_{CO}$ to $0.03T_{CO}$, 
and the difference between the ``bare'' rate and the rate adjusted by the Lerch transcendent 
formula of Refs.\cite{bradley_bose-einstein_2008,blakie_dynamics_2008,BradleyLowDimensional2015}. 
In practice, the adjustment to the 
``bare'' rate is small (of order 1) 
for these parameters. For ${}^{7}{\rm Li}$ the range is 
$3.4\times 10^{-6}\leq\gamma\leq 1.2\times 10^{-5}$. This suggests 
suppression of the resonances by thermal damping would not be achieved. However, 
this conclusion rests on the model we employed and the assumption that the scenario 
is sufficiently near-equilibrium for the {\it a priori} prediction of $\gamma$ to be relevant. As mentioned 
above, effective $\gamma$ values that would be sufficient to suppress some resonances have 
been used in previous SPGPE modelling of non-equilibrium experiments. Considerations 
outside our model that might affect this conclusion include 3D effects in the 
quasi-condensate not captured by our quasi-1D model, energy-damping terms due to 
scattering not captured by our simple-growth SPGPE model \cite{RooneyStochastic2012}, 
and the previously mentioned possibility of the resonance driving dynamics of the 
thermal cloud not captured in the SPGPE model. These all might alter the effective $\gamma$.

\section{Conclusion}

We have shown using numerical modelling that an ultracold gas of atoms in two hyperfine 
states can act as an analogue to a relativistic system undergoing a first order phase 
transition, such as might have occurred in the very early universe. In particular, we 
have investigated whether the parametric resonance effects which previously were 
noted to be problematic for the systems based on an oscillatory potential might be 
overcome by dissipative effects. Our results suggest this is in fact unlikely for 
reasonable experimental parameters, based on the {\it a priori} predicted damping rates. 
However, uncertainty remains over whether the {\it a priori} estimate of the damping
rate would apply to a non-equilibrium experiment of this type, and over effects not 
captured by our simple growth SPGPE model with a static thermal cloud. Experimental 
measurements of the effective $\gamma$ under conditions similar to those described above, performed for example by measuring experimental rates of growth towards equilibrium and comparing these to an SPGPE model \cite{weiler_spontaneous_2008,ota_collisionless_2018,liu_dynamical_2018}, would be worthwhile to determine 
this conclusively. Further theoretical investigation of the oscillating potential setup using a 
description that includes thermal cloud dynamics would be an interesting avenue for further work.

However, our results from numerical modelling are in good agreement with a purely 
theoretical treatment of the effect of thermal damping on the parametric resonance 
bands. Having the system under such good theoretical control should be useful in 
designing future experiments. One feature which we have not been able to reproduce, 
and deserves further investigation, is the possibility of parametric amplification of the thermal noise.

A particularly interesting development of the present system would be to extend it to 
two or three dimensions. This would allow a far richer picture of interacting bubbles than in one 
dimension. Two dimensions would likely be optimal as the bubbles could be imaged 
in an experiment without being obscured by the surrounding gas. The theoretical treatment extends easily 
to two or three dimensions, but the main complication is the presence of boundaries. 
Other treatments of bubble nucleation in two dimensions have shown that the boundary 
can act as a site of bubble nucleation \cite{BillamSimulating2019}. Both this, and extensions 
to metastable states of systems with a larger number of components, are avenues 
we hope to investigate in future work.

\section*{Acknowledgements}
This work was supported in part by the UK Engineering and Physical Sciences Research Council 
[grant EP/R021074/1], the Science and Technology Facilities Council (STFC) [grant ST/T000708/1]
and the UK Quantum Technologies for Fundamental Physics programme [grant ST/T00584X/1]. 
KB is supported by an STFC studentship. This research made use of the Rocket High Performance 
Computing service at Newcastle University. 

\appendix

\section{Parametric amplification}

In this appendix we complete the theoretical analysis of resonant effects by including the noise terms.
Noise in time-periodic stochastic systems drives the phenomenon of parametric
amplification, which occurs close to a resonance 
band even when the parametric resonance is damped out. This effect amplifies the 
effectiveness of the noise, increases the fluctuations in the phase and could potentially
trigger early bubble nucleation.

The response of the linearised system Eqs. (\ref{eq1}) and (\ref{eq2}) to the noise terms can be written in terms of a 
matrix Green function $G_{ij}(t,t')$, for example
\begin{equation}
\varphi(t)=\int_0^t dt'\,\left\{
G_{\varphi\varphi}(t,t')\eta_\varphi(t')+G_{\varphi\sigma}(t,t')\eta_\sigma(t')\right\}
\end{equation}
Suppose the eigenvalues of the averaged system (\ref{eqA}) and (\ref{eqB}) are $\lambda_-$ and $\lambda_+$.
Both of these are negative in the damped regime, and we choose $|\lambda_+|<|\lambda_-|$. The
approximate solutions Eq. (\ref{resphi}) for $\varphi$ are
\begin{align}
\varphi_+&=e^{\lambda_+t}\left(\cos\omega t/2-\sin\omega t/2\right),\\
\varphi_-&=e^{\lambda_-t}\left(\cos\omega t/2+\sin\omega t/2\right).
\end{align}
Using standard Green function methods, the $\varphi\sigma$ component of the Green function to 
leading order in $\epsilon$ is
\begin{equation}
G_{\varphi\sigma}(t,t')=
{\varphi_+(t')\varphi_-(t)-\varphi_-(t')\varphi_+(t)\over 
\sigma_+(t')\varphi_-(t')-\sigma_-(t')\varphi_+(t')}\theta(t-t'),\label{gmod}
\end{equation}
where $\theta(t)$ is the Heaviside function and $\sigma_\pm\approx -\dot\varphi_\pm/a$
from Eq. (\ref{eq1}).

The correlation function for the phase driven by the $\eta_\sigma$ noise becomes
\begin{equation}
\langle\varphi(k,t)\varphi^*(k',t)\rangle=2\gamma T\int_0^t dt'\,G_{\varphi\sigma}(t,t')^2 \delta_{kk'},
\end{equation}
where the dependence on wavenumber is implicit in the modes (\ref{gmod}).
The rapidly oscillating terms average out, and we are left with
\begin{equation}
\langle\varphi(k,t)\varphi^*(k',t)\rangle\approx{a^2\gamma T\over \omega^2}
\left({1\over|\lambda_+|}+{1\over|\lambda_-|}\right)\delta_{kk'}.
\end{equation}
At the centre of the resonance band, $\lambda_+$ is given by Eq. (\ref{Gamma}), and
\begin{equation}
\langle\varphi(k,t)\varphi^*(k',t)\rangle\approx{a\gamma T\over 4b|4\delta-(2+k^2)\gamma|}\delta_{kk'}.
\end{equation}
Thus the variance in the phase is enhanced when the denominator, $|\lambda_+|$, is small.
However, the variance calculated here is an equilibrium value. In practice, preparation 
of the system in a non-equilibrium state can lead to a smaller variance initially,
and the fluctuations may take a  significant time to reach the equilibrium value, 
at a rate depending on the eigenvalue $\lambda_-$. This may explain why
our numerical simulations show no evidence of enhanced bubble nucleation
in the small $|\lambda_+|$ regime.

\bibliography{paper}

\end{document}